\documentclass[aps,prb,twocolumn,groupedaddress,floatfix]{revtex4-1}

\usepackage{graphicx}
\usepackage{dcolumn}
\usepackage{bm}

\begin{document}

\title{Anisotropic effects and phonon induced spin relaxation in gate-controlled semiconductor quantum dots}
\author{Sanjay Prabhakar,$^{1}$ Shohini Ghose,$^{2}$  Roderick Melnik$^{1,3}$ and Luis L. Bonilla$^{3}$}
\affiliation{
$^1$M\,$^2$NeT Laboratory, Wilfrid Laurier University, Waterloo, ON, N2L 3C5 Canada\\
$^2$Department of Physics and Computer Science, Wilfrid Laurier University, Waterloo, ON, N2L 3C5 Canada\\
$^3$Gregorio Millan Institute, Universidad Carlos III de Madrid, 28911, Leganes, Spain}

\date{January 26, 2012}

\begin{abstract}
In this paper, a detailed analysis of anisotropic effects on the phonon induced spin relaxation rate
in III-V semiconductor quantum dots (QDs) is carried out. We show that the accidental degeneracy
due to level crossing between the first and second excited states of opposite electron spin states in
both isotropic and anisotropic QDs can be manipulated with the application of externally applied
gate potentials. In particular, anisotropic gate potentials enhance the phonon mediated spin-flip
rate and reduce the cusp-like structure to lower magnetic fields, in addition to the lower QDs radii
in III-V semiconductor QDs. In InAs QDs, only the Rashba spin-orbit coupling contributes to the
phonon induced spin relaxation rate. However, for GaAs QDs, the Rashba spin-orbit coupling has
a contribution near the accidental degeneracy point and the Dresselhaus spin-orbit coupling has
a contribution below and above the accidental degeneracy point in the manipulation of phonon
induced spin relaxation rates in QDs.
\end{abstract}



\maketitle

\section{Introduction}
The study of electron spin states in zero dimensional semiconductor nanostructures such as QDs is important
for the development of next generation electronic devices such as spin transistors, spin filters, spin memory devices
and quantum logic gates.~\cite{loss98,awschalom02,hanson05, kroutvar04,elzerman04,glazov10}
The electron spin states in QDs are brought in resonance or out of resonance by
applying suitable gate potentials in order to read out the spin states.~\cite{bandyopadhyay00,fujisawa01} Progress in nanotechnology has made it
possible to fabricate gated quantum dots with desirable optoelectronic and spin properties.~\cite{folk01,fujisawa01,fujisawa02,hanson03} Very recently, it
was shown that the electron spin states in gated quantum dots can be measured in the presence of magnetic fields
along arbitrary directions.~\cite{takahashi10,deacon11,kanai11,marquardt11,kanai10}

A critical ingredient for the design of robust spintronic devices is the accurate estimation of the spin relaxation
rate. Recent studies by authors in Refs.~\onlinecite{elzerman04,kroutvar04}  have measured long spin relaxation times of 0.85 ms
in GaAs QDs by pulsed relaxation rate measurements and 20 ms in InGaAs QDs by optical orientation measurements.
These experimental studies in QDs confirm that the manipulation of spin-flip rate by spin-orbit coupling
with respect to the environment is important for the design of robust spintronics logic devices.~\cite{golovach04,khaetskii00,prabhakar10} The
spin-orbit coupling is mainly dominated by the Rashba~\cite{bychkov84} and the linear Dresselhaus~\cite{dresselhaus55} terms in solid state QDs.
The Rashba spin-orbit coupling arises from structural inversion asymmetry along the growth direction and the
Dresselhaus spin-orbit coupling arises from the bulk inversion asymmetry of the crystal lattice.~\cite{bulaev05,bulaev05a,sousa03} Recently,
electric and magnetic fields tunability of the electron spin states in gated III-V semiconductor QDs was manipulated
through Rashba and Dresselhaus spin-orbit couplings.~\cite{sousa03,prabhakar09,pryor06,pryor07,nowak11}

Anisotropic effects induced in the orbital angular momentum in QDs suppresses the Land$\acute{e}$ $g$-factor towards
bulk crystal.~\cite{prabhakar09,prabhakar11} $g$-factor can be manipulated through strong Rashba spin-orbit coupling in InAs QDs~\cite{prabhakar11} and
trough strong Dresselhaus spin-orbit coupling in GaAs QDs.~\cite{prabhakar09} Large anisotropy effects of the spin-orbit interaction
in self-assembled InAs QDs have been recently studied experimentally in Ref.~\onlinecite{takahashi10}. In this paper, we study
the phonon induced spin-flip rate of electron spin states in both isotropic and anisotropic QDs. Our studies show
that the Rashba spin-orbit coupling has an appreciable contribution to the spin-flip rate in InAs QDs. However,
the Rashba spin-orbit coupling has a contribution to the spin-flip rate in GaAs QDs near the level crossing
point and the Dresselhaus spin-orbit coupling elsewhere. Anisotropic gate potentials, playing an important role in
the spin-flip rate, can be used to manipulate the accidental degeneracy due to level crossing and avoided anticrossing
between the electron spin states $|0, 0,->$ and $|0, 1,+ >$. In this paper, we show that the anisotropic
gate potentials cause also a quenching effect in the orbital angular momentum that enhances the phonon mediated
spin-flip rate and reduces its cusp-like structure to lower magnetic fields, in addition to lower QDs radii.

The paper is organized as follows: In section II, we develop a theoretical model and find an analytical expression
for the energy spectrum of electron spin states for anisotropic QDs. In section III, we find the spin relaxation
rate of electron spin states for anisotropic and isotropic QDs. In section IV, we show that the cusplike
structure in spin-flip rate due to accidental degeneracy points in QDs can be manipulated to lower magnetic
fields, in addition to lower QDs radii, with the application of anisotropic gate potentials. Different mechanisms
of spin-orbit interactions such as Rashba vs. Dresselhaus couplings are also discussed in this section. Finally, in
section V, we summarize our results.

\begin{figure}
\includegraphics[width=8.5cm,height=6cm]{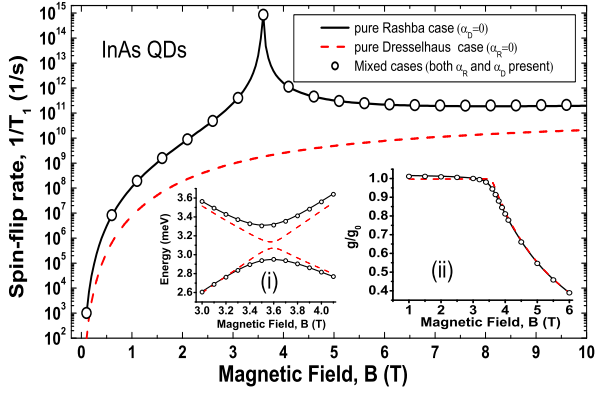}
\caption{\label{fig1}
(Color online) Contributions of Rashba  and Dresselhaus   spin-orbit couplings on the phonon induced spin-flip rate vs magnetic fields in InAs QDs.  Inset plots  shows the energy difference vs. magnetic fields near the level crossing point and  the $g$-factor vs magnetic fields.  Here we choose $E=10^5$ V/cm, $\ell_0=20$ nm and $a=b=1$. The material constants for InAs QDs are chosen from Refs.~\onlinecite{sousa03,cardona88} as $g_0=-15$, $m=0.0239$, $\gamma_R=110~ \mathrm{{\AA}^2}$, $\gamma_D=130~\mathrm{eV{\AA}^3}$, $eh_{14}=0.54\times 10^{-5} ~\mathrm{erg/cm}$, $s_l=4.2\times 10^{5} ~\mathrm{cm/s}$, $s_t=2.35\times 10^{5}~\mathrm{cm/s}$ and $\rho=5.6670 ~\mathrm{g/cm^3}$.
}
\end{figure}
\begin{figure*}
\includegraphics[width=18cm,height=6cm]{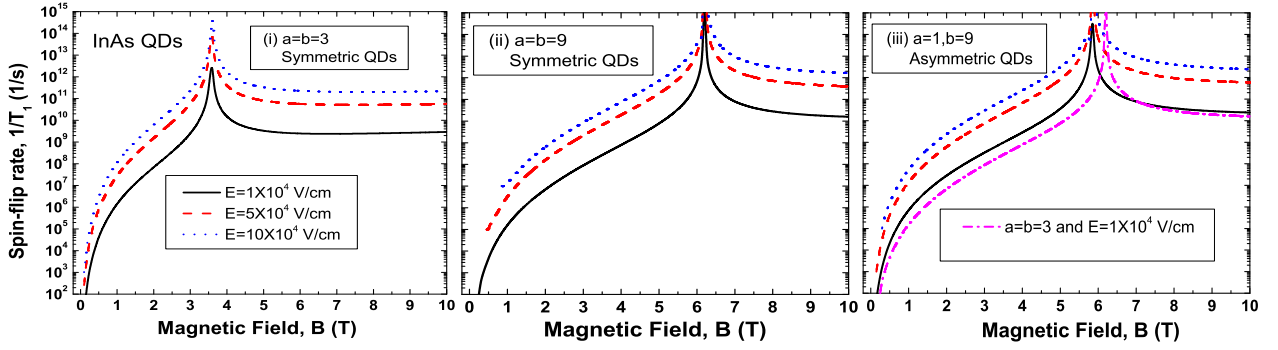}
\caption{\label{fig2}
(Color online) Spin relaxation rate ($1/T_1$) vs magnetic fields  between the states $|00+>$ and $|00->$ in InAs QDs.
Here we chose $\ell_0=20 nm$  (the QD radius). As a reference in Fig.~\ref{fig2}(iii), dashed-dotted line represents the spin-flip rate for symmetric QDs with $a = b = 3$. We see that the anisotropic potential enhances the spin-flip rate and reduces the level crossing point to lower magnetic field.
}
\end{figure*}
\begin{figure}
\includegraphics[width=8.5cm,height=6cm]{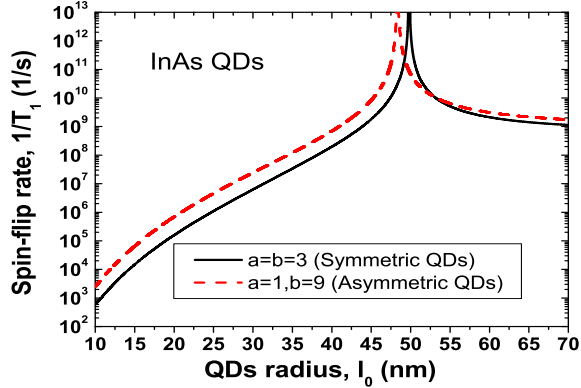}
\caption{\label{fig3}
(Color online) Spin relaxation rate ($1/T_1$) vs QDs radius between the states $|00+>$ and $|00->$ in InAs QDs.
Here, we choose $B = 1 T$ and $E = 10^4 V/cm$ (magnetic and electric fields, respectively). We see that the anisotropic
potential reduces the level crossing point to lower QDs radius.
}
\end{figure}
\begin{figure}
\includegraphics[width=8.5cm,height=6cm]{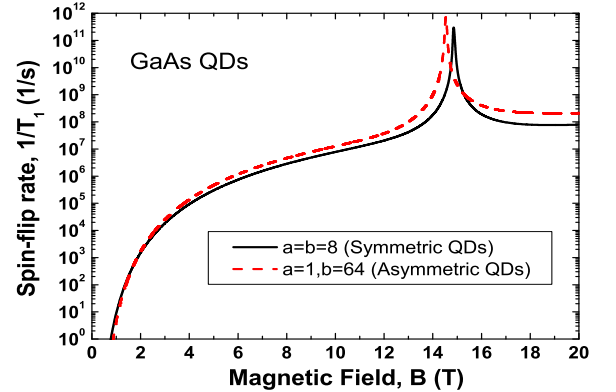}
\caption{\label{fig4}
(Color online) Spin relaxation rate ($1/T_1$) vs magnetic fields between the states $|00+>$ and $|00->$ in GaAs QDs.
Here we chose, QDs radius, $\ell_0=32 nm$  and electric field, $E = 10^4 V/cm$. Again, we see that the level crossing point reduces to the lower magnetic fields. For GaAs QDs, we use the material constants from Refs.~\onlinecite{sousa03,cardona88} as  $g_0=-0.44$, $m=0.067$, $\gamma_R=4.4 ~\mathrm{{\AA}^2}$, $\gamma_D=26~\mathrm{eV{\AA}^3}$, $eh_{14}=2.34\times 10^{-5} ~\mathrm{erg/cm}$, $s_l=5.14\times 10^{5} ~\mathrm{cm/s}$, $s_t=3.03\times 10^{5} ~\mathrm{cm/s}$ and $\rho=5.3176 ~\mathrm{g/cm^3}$
}
\end{figure}
\begin{figure}
\includegraphics[width=8.5cm,height=6cm]{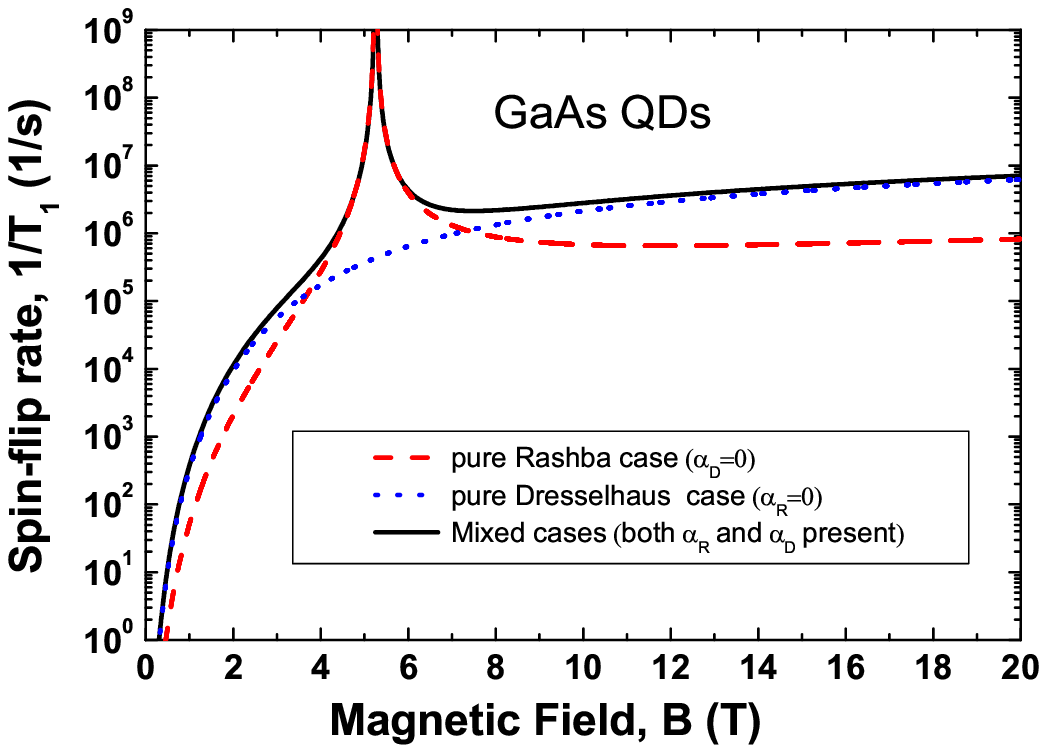}
\caption{\label{fig5}
(Color online)
Contributions of Rashba and Dresselhaus spin-orbit couplings on the phonon induced spin-flip rate
as a function of magnetic fields in GaAs QDs. Here we choose $E = 10^5 V/cm$, $\ell_0=32 nm$  and $a=b=1$. We see that only Rashba spin-orbit coupling gives the cusp-like structure in the spin-flip rate.
}
\end{figure}

\section{Theoretical Model}\label{theoretical-model}
We consider 2D anisotropic III-V semiconductor QDs in the presence of a magnetic field along the growth
direction. The total Hamiltonian of an electron in anisotropic QDs including spin-orbit interaction can be
written as~\cite{sousa03,khaetskii00,bulaev05}
\begin{equation}
H = H_{xy} +  H_{so},
\label{total}
\end{equation}
where the Hamiltonian $H_{so}$ is associated with the Rashba-Dresselhaus spin-orbit couplings and $H_{xy}$ is the
Hamiltonian of the electron in anisotropic QDs. $H_{xy}$ can be written as
\begin{equation}
H_{xy} = {\frac {\vec{P}^2}{2m}} + {\frac{1}{2}} m \omega_o^2
(a x^2 + b y^2) + {\frac 1 2} g_o \mu_B \sigma_z B,
\label{hxy}
\end{equation}
where $\vec{P} = \vec{p} + e \vec{A}$ is  the kinetic momentum operator, $\vec{p} = -i\hbar (\partial_x,\partial_y,0)$ is the canonical momentum operator, $\vec{A}$ is the vector potential in the asymmetric gauge, $m$ is the effective mass of the electron in the conduction band,  $\mu_B$ is the Bohr magneton, $\vec{\sigma}=\left(\sigma_x,\sigma_y,\sigma_z\right)$ are the Pauli spin matrices, $\omega_0=\frac{\hbar}{m\ell_0^2}$ is the parabolic confining potential and $\ell_0$ is the radius of the QDs.

To find the energy spectrum of Hamiltonian~(\ref{hxy}), it is convenient to introduce  the canonical transformation of position and momentum operator as~\cite{schuh85,galkin04}
\begin{eqnarray}
x_1=a^{1/4} x,\quad P_1=\frac{1}{a^{1/4}}P_x, \label{x-1}\\
x_2=b^{1/4} y,\quad P_2=\frac{1}{b^{1/4}} P_y. \label{x-2}
\end{eqnarray}
Also, the assymetric Gauge potential can be written as
\begin{equation}
A_x=-x_2B\left(\frac{b^{1/4}}{\sqrt a+\sqrt b}\right),A_y=x_1B\left(\frac{a^{1/4}}{\sqrt a+\sqrt b}\right).\label{A-x}
\end{equation}
By substituting Eqs. (\ref{x-1},~\ref{x-2},~\ref{A-x}) into Eq.~(\ref{hxy}), we get the Hamiltonian in the form:
\begin{equation}
h=P_1^2 + x_1^2 + \Im\left(P_2^2+x_2^2\right)+\wp\left(x_1P_2-x_2P_1\right) +  {\frac 1 2} g_o \mu_B \sigma_z B,
\label{h}
\end{equation}
where $h=\frac{2m}{\sqrt a}H_{xy}$, $\Im=\sqrt{\frac{b}{a}}$, $\wp=\frac{2\omega_c\left(b/a\right)^{1/4}}{\left[\omega_c^2+\omega_0^2\left(\sqrt a+\sqrt b\right)^2\right]^{1/2}}$. Also we use the relation $m\omega_0\gamma=1$, where $\gamma^2=1+ \frac{\left(\omega_c/\omega_0\right)^2}{\left(\sqrt a+ \sqrt b\right)^2}  $ and $\omega_c=\frac{eB}{m}$ is the cyclotron frequency.

The energy  spectrum of  Hamiltonian (\ref{h}) can be found as follows.
First, we need  to find the canonical transformation $U$ of the four-dimensional phase space, $P^t\equiv\left(P_1,P_2,x_1,x_2\right)$   which diagonalizes  the
quadratic form of the Hamiltonian (\ref{h}). To be more specific,  Hamiltonian (\ref{h}) without Zeeman spin splitting energy can be written as
\begin{equation}
h=P^t\mathbf{M}P,~
\mathbf{M}=
\left(\begin{array}{cccc}
1 & 0 & 0 & -\wp/2 \\
0 & \Im & \wp/2 & 0 \\
0 & \wp/2 & 1 & 0 \\
-\wp/2 & 0 & 0 & \Im \\
\end{array}\right),
\label{hxy-c}
\end{equation}
where $t$ represents  the transpose of a vector. The orthogonal unitary matrix $\mathbf{U}$ which exactly diagonalizes the matrix $\mathbf{M}$ can be written as,
\begin{equation}
\mathbf{U}=\frac{1}{\left(s_+-s_-\right)}
\left(\begin{array}{cccc}
1 & 1 & -s_- & -s_+ \\
1 & -1 & s_+ & -s_- \\
s_- & s_+ & 1 & 1 \\
-s_+ & s_- & 1 & -1 \\
\end{array}\right),
\label{U}
\end{equation}
where $\wp s_{\pm}\equiv \Im-1\pm d$ and
\begin{eqnarray}
s_{\pm}=\frac{\omega_+}{\omega_c \left(\frac{b}{a}\right)^{\frac{1}{4}}}\left[\sqrt \frac{b}{a}-1 \pm \left[\frac{\omega_c^2\sqrt \frac{b}{a}}{\omega_+^2}+\left(1-\sqrt \frac{b}{a}\right)^2\right]^{\frac{1}{2}}\right],\label{s-pm}~~~\\
\omega_{\pm}=\frac{1}{2}\left[\omega_c^2+\omega_0^2\left(\sqrt a\pm\sqrt b\right)^2\right]^{1/2},~~~~\label{omega-pm}
\end{eqnarray}
\begin{equation}
d=\left[\frac{4\omega_c^2\sqrt {\frac{b}{a}}}{\omega_c^2+\omega_0^2\left(\sqrt a+\sqrt b\right)^2}+\left(1-\sqrt {\frac{b}{a}}\right)^2\right]^{\frac{1}{2}}.
\label{d}
\end{equation}
In the form  of rotated operators   $P'=\mathbf{U}P$, the Hamiltonian (\ref{h}) can be written as
\begin{eqnarray}
h=\frac{1}{2}(\wp s_-+2)\left(p_x^{'2}+x^{'2}\right)+\frac{1}{2}(\wp s_++2)\left(p_y^{'2}+y^{'2}\right).~~~~~~
\label{h-a}
\end{eqnarray}
The above Hamiltonian represents  the  superposition of two independent harmonic oscillators. The  energy spectrum of $H_{xy}$ can be written as
\begin{equation}
\varepsilon_{n_+n_-}=\left(n_++n_-+1\right)\hbar\omega_++\left(n_+-n_-\right)\hbar\omega_-+\frac{1}{2}g_0\mu_B\sigma_zB, \label{epsilon-a}\\
\end{equation}
where $n_{\pm}=a_{\pm}^\dagger a_{\pm}$ are the number operators. Here,  $a_{\pm}$ and $a_{\pm}^\dagger$ are usual annihilation (``lowering'') and creation (``raising'') operators. In Eq.~\ref{epsilon-a}, we included Zeeman spin splitting energy.


The Hamiltonian associated with the Rashba and linear Dresselhaus spin-orbit couplings  can be written as~\cite{bychkov84,dresselhaus55,sousa03}
\begin{equation}
H_{so} =\frac{\alpha_R}{\hbar}\left(\sigma_x P_y - \sigma_y P_x\right)+ \frac{\alpha_D}{\hbar}\left(-\sigma_x P_x + \sigma_y P_y\right),\label{rashba-dresselhaus}
\end{equation}
where the strengths of the Rashba and Dresselhaus spin-orbit couplings are characterized by the parameters $\alpha_R$ and $\alpha_D$. They are given by
\begin{equation}
\alpha_R=\gamma_ReE,~~~~~\alpha_D=0.78\gamma_D\left(\frac{2me}{\hbar^2}\right)^{2/3}E^{2/3},\label{coefficient-R-D}
\end{equation}
where  $\gamma_R$ and   $\gamma_D$ are the Rashba and Dresselhaus coefficients.

The energy spectrum of the Hamiltonian associated with the Rashba and Dresselhaus spin-orbit couplings can be written as
\begin{widetext}
\begin{eqnarray}
H_{so}&=&\alpha_R \left(1+i\right)\left[
b^{1/4}\kappa_+\left(s_+-i\right)a_++b^{1/4}\kappa_+\left(s_-+i\right)a_-+a^{1/4}\eta_-\left(i-s_-\right)a_++a^{1/4}\eta_-\left(i+s_+\right)a_-
\right]\nonumber\\
&&+\alpha_D
\left(1+i\right)\left[a^{1/4}\kappa_-\left(i-s_-\right)a_++a^{1/4}\kappa_-\left(i+s_+\right)a_-+b^{1/4}\eta_+\left(-i+s_+\right)a_++b^{1/4}\eta_+\left(i+s_-\right)a_-\right]+H.c.,~~~~~~
\label{H-R}
\end{eqnarray}
\end{widetext}
where,
\begin{eqnarray}
\kappa_{\pm}=\frac{1}{2\left(s_+-s_-\right)}\left\{\frac{1}{\ell}\sigma_x\pm i\frac{eB\ell}{\hbar}\left(\frac{1}{\sqrt{a}+\sqrt{b}}\right)\sigma_y\right\},~~~~\\
\eta_{\pm}=\frac{1}{2\left(s_+-s_-\right)}\left\{\frac{1}{\ell}\sigma_y\pm i\frac{eB\ell}{\hbar}\left(\frac{1}{\sqrt{a}+\sqrt{b}}\right)\sigma_x\right\},~~~~
\end{eqnarray}
where H.c. represents the Hermitian conjugate, $\ell=\sqrt{\hbar/m\Omega}$ is the hybrid orbital length and $\Omega=\sqrt{\omega_0^2+\omega_c^2/4}$. It is clear that the spin-orbit Hamiltonian
and the Zeeman spin splitting energy in both isotropic
and anisotropic QDs obey a selection rule in which the
orbital angular momentum can change by one quantum.

At low electric fields and small QDs radii, we treat
the Hamiltonian associated with the Rashba and linear
Dresselhaus spin-orbit couplings as a perturbation. Using
second order perturbation theory, the energy spectrum
of the electron spin states in QDs is given by
\begin{eqnarray}
\varepsilon_{0,0,+}=\hbar \varpi_+ - \frac{\alpha_R^2\xi_++\alpha_D^2\varsigma_+}{\hbar \omega_x-\Delta}-\frac{\alpha_R^2\varsigma_-+\alpha_D^2\xi_-}{\hbar \omega_y-\Delta},\\
\varepsilon_{0,0,-}=\hbar \varpi_- - \frac{\alpha_R^2\varsigma_++\alpha_D^2\xi_+}{\hbar \omega_x+\Delta}-\frac{\alpha_R^2\xi_-+\alpha_D^2\varsigma_-}{\hbar \omega_y+\Delta},
\end{eqnarray}
where, $\varpi_{\pm}=\omega_+\pm\omega_z/2$, $\omega_z=\Delta/\hbar$ is the Zeeman frequency, $\Delta=g_0\mu_BB$, $\omega_x=\omega_++\omega_-$, and $\omega_y=\omega_+-\omega_-$. Also,
\begin{eqnarray}
\xi_{\pm}=\frac{1}{2(s_+-s_-)}\left\{\pm\frac{1}{s_{\pm}}\alpha^2_{\pm}+2\alpha_{\pm}\beta_{\pm} \mp\frac{1}{s_{\mp}}\beta^2_{\pm}\right\},\\
\varsigma_{\pm}=\frac{1}{2(s_+-s_-)}\left\{\pm\frac{1}{s_{\pm}}\alpha^2_{\mp}-2\alpha_{\mp}\beta_{\mp} \mp\frac{1}{s_{\mp}}\beta^2_{\mp}\right\},\\
\alpha_{\pm}=a^{1/4}\left\{\frac{1}{\ell}\pm \frac{eB\ell}{\hbar}\frac{1}{\left(\sqrt{a}+\sqrt{b}\right)}\right\},\\
\beta_{\pm}=b^{1/4}\left\{\frac{1}{\ell}\pm \frac{eB\ell}{\hbar}\frac{1}{\left(\sqrt{a}+\sqrt{b}\right)}\right\}.
\end{eqnarray}

\section{Phonon induced spin relaxation}\label{spin-relaxation}
We now turn to the calculation of the phonon induced
spin relaxation rate between two lowest energy states in
QDs. Following Ref.~\onlinecite{prabhakar12}, the interaction between electron
and piezo-phonon can be written as~\cite{khaetskii00,khaetskii01}
\begin{equation}
u^{\mathbf{q}\alpha}_{ph}\left(\mathbf{r},t\right)=\sqrt{\frac{\hbar}{2\rho V \omega_{\mathbf{q}\alpha}}} e^{i\left(\mathbf{q\cdot r} -\omega_{q\alpha} t\right)e A_{\mathbf{q}\alpha}b^{\dag}_{\mathbf{q}\alpha}} + H.c.
\label{u}
\end{equation}
Here, $\rho$ is the crystal mass density, $V$ is the volume of the QDs, $b^{\dag}_{\mathbf{q}\alpha}$ creates an acoustic phonon with wave vector $\mathbf{q}$ and polarization $\hat{e}_\alpha$, where $\alpha=l,t_1,t_2$ are chosen as one longitudinal  and two transverse modes of the induced phonon  in the dots. Also,  $A_{\mathbf{q}\alpha}=\hat{q}_i\hat{q}_k e\beta_{ijk} e^j_{\mathbf{q}\alpha}$ is the amplitude of the electric field created by phonon strain, where $\hat{\mathbf{q}}=\mathbf{q}/q$ and $e\beta_{ijk}=eh_{14}$ for $i\neq k, i\neq j, j\neq k$. The polarization directions of the induced phonon are $\hat{e}_l=\left(\sin\theta \cos\phi, \sin\theta \sin\phi, \cos\theta \right)$, $\hat{e}_{t_1}=\left(\cos\theta \cos\phi, \cos\theta \sin\phi, -\sin\theta \right)$ and $\hat{e}_{t_2}=\left(-\sin\phi, \cos\phi, 0 \right)$. Based on the Fermi Golden Rule, the phonon induced spin transition rate in the QDs is given by~\cite{sousa03,khaetskii01}
\begin{equation}
\frac{1}{T_1}=\frac{2\pi}{\hbar}\int \frac{d^3\mathbf{q}}{\left(2\pi\right)^3}\sum_{\alpha=l,t}\arrowvert M\left(\mathbf{q}\alpha\right)\arrowvert^2\delta\left(\hbar s_\alpha \mathbf{q}-\varepsilon_{0,0,-}+\varepsilon_{0,0,+}\right),
\label{1-T1}
\end{equation}
where  $s_l$,$s_t$ are the longitudinal and transverse acoustic phonon velocities in QDs. The matrix element $M\left(\mathbf{q}\alpha\right)$ for the spin-flip rate between the Zeeman sublevels with the emission of phonon $\mathbf{q}\alpha$ has been calculated perturbatively.~\cite{khaetskii01,stano06} As a result, we have:
\begin{equation}
\frac{1}{T_1}=c\left(|M_x|^2+|M_y|^2\right),
\label{1-T1-a}
\end{equation}
where,
\begin{widetext}
\begin{eqnarray}
c=\frac{\left(eh_{14}\right)^2\left(g\mu_BB\right)^3}{35\pi \hbar^4\rho}\left(\frac{1}{s^5_l}+\frac{4}{3}\frac{1}{s^5_t}\right),~~~~~\label{c}\\
M_x=\frac{\left(is_-+1\right)\Xi_1\left(\hbar\omega_x+\Delta\right)+\left(-is_-+1\right)\Xi_3\left(\hbar\omega_x-\Delta\right)}
{a^{1/4}\left[\left(\hbar\omega_x\right)^2-\Delta^2\right]}
+\frac{\left(-is_++1\right)\Xi_2\left(\hbar\omega_y+\Delta\right)+\left(is_++1\right)\Xi_4\left(\hbar\omega_y-\Delta\right)}
{a^{1/4}\left[\left(\hbar\omega_y\right)^2-\Delta^2\right]},~~~\label{M-x}\\
M_y=\frac{\left(is_++1\right)\Xi_1\left(\hbar\omega_x+\Delta\right)+\left(-is_++1\right)\Xi_3\left(\hbar\omega_x-\Delta\right)}
{b^{1/4}\left[\left(\hbar\omega_x\right)^2-\Delta^2\right]}
+\frac{\left(is_--1\right)\Xi_2\left(\hbar\omega_y+\Delta\right)+\left(-is_--1\right)\Xi_4\left(\hbar\omega_y-\Delta\right)}
{b^{1/4}\left[\left(\hbar\omega_y\right)^2-\Delta^2\right]},~~~\label{M-y}\\
\Xi_1=\frac{\ell}{2\left(s_+-s_-\right)^2}\left[\alpha_R\left\{\left(s_++i\right)\beta_++\left(1-is_-\right)\alpha_+\right\}
+\alpha_D\left\{\left(-s_--i\right)\alpha_-+\left(-1+is_+\right)\beta_-\right\}\right],~~~~\label{Xi-1}\\
\Xi_2=\frac{\ell}{2\left(s_+-s_-\right)^2}\left[\alpha_R\left\{\left(s_--i\right)\beta_++\left(1+is_+\right)\alpha_+\right\}
+\alpha_D\left\{\left(s_+-i\right)\alpha_-+\left(1+is_-\right)\beta_-\right\}\right],~~~~\label{Xi-2}\\
\Xi_3=\frac{\ell}{2\left(s_+-s_-\right)^2}\left[\alpha_R\left\{\left(s_+-i\right)\beta_-+\left(-1-is_-\right)\alpha_-\right\}
+\alpha_D\left\{\left(-s_-+i\right)\alpha_++\left(1+is_+\right)\beta_+\right\}\right],~~~~\label{Xi-3}\\
\Xi_4=\frac{\ell}{2\left(s_+-s_-\right)^2}\left[\alpha_R\left\{\left(s_-+i\right)\beta_-+\left(-1+is_+\right)\alpha_-\right\}
+\alpha_D\left\{\left(s_++i\right)\alpha_++\left(-1+is_-\right)\beta_+\right\}\right].~~~~\label{Xi-4}
\end{eqnarray}
\end{widetext}
In the above expression, we use $c=c_l I_{xl}+2c_tI_{xt}$ where $c_\alpha=\frac{q^2e^2}{\left(2\pi\right)^2\hbar^2s_\alpha}|\varepsilon_{q\alpha}|^2$, $|\varepsilon_{q\alpha}|^2=\frac{q^2\hbar}{2\rho\omega_{q\alpha}}$ and $q=\frac{g\mu_BB}{\hbar s_\alpha}$. Also, $g=\frac{\varepsilon_{0,0,-}-\varepsilon_{0,0,+}}{\mu_BB}$ is the Land$\acute{e}$ $g$-factor. Also, for longitudinal phonon modes $I_{xl}=I_{yl}=\frac{8\pi}{35}h^2_{14}$ and $I_{xyl}=0$. For transverse phonon modes, $I_{xt}=I_{yt}=\frac{16\pi}{105}h^2_{14}$ and $I_{xyt}=0$.

For isotropic QDs ($a=b=1$, $s_+=1$ and  $s_-=-1$), the spin relaxation rate is given by
\begin{equation}
\frac{1}{T_1}=\frac{\left(eh_{14}\right)^2\left(g\mu_BB\right)^3}{35\pi \hbar^4\rho}\left(\frac{1}{s^5_l}+\frac{4}{3}\frac{1}{s^5_t}\right)\left(|M_R|^2+|M_D|^2\right),
\label{1-T1-1}
\end{equation}
where $M_R$ and $M_D$ are the coefficients of matrix element
associated to the Rashba and Dresselhaus spin-orbit coupling
in QDs and is given by
\begin{eqnarray}
M_R=\frac{\alpha_R}{\sqrt 2 \hbar \Omega}\left[\frac{1}{1-\frac{\Delta}{\hbar\left(\Omega+\frac{\omega_c}{2}\right)}}-
\frac{1}{1+\frac{\Delta}{\hbar\left(\Omega-\frac{\omega_c}{2}\right)}}\right],\label{M-R}\\
M_D=\frac{\alpha_D}{\sqrt 2 \hbar \Omega}\left[\frac{1}{1+\frac{\Delta}{\hbar\left(\Omega+\frac{\omega_c}{2}\right)}}-
\frac{1}{1-\frac{\Delta}{\hbar\left(\Omega-\frac{\omega_c}{2}\right)}}\right].\label{M-D}
\end{eqnarray}
Since $\Delta = g_0\mu_BB$ is negative for GaAs and InAs QDs, it means the degeneracy only appears in the Rashba case
(see 2nd term of Eq.~\ref{M-R}) and is absent in the Dresselhaus case. Similarly, the degeneracy only appears in the $g$-factor for the Rashba spin-orbit coupling.~\cite{prabhakar11}  The degeneracy in the Rashba case induces the
level crossing point and cusp-like structure in the spin-flip rate in QDs. By considering second power of $\Delta$, the
spin relaxation rate for isotropic QDs is given by
\begin{eqnarray}
\frac{1}{T_1}&=&\frac{\left(eh_{14}\right)^2\left(g\mu_BB\right)^3}{35\pi \hbar^4\rho}\left(\frac{1}{s^5_l}+\frac{4}{3}\frac{1}{s^5_t}\right)\frac{2\Delta^2}{\hbar^4\Omega^4}\left(\alpha_R^2+\alpha_D^2\right)\nonumber\\
&&\left[1+2\left(\frac{\omega_c}{2\Omega}\right)^2+3\left(\frac{\omega_c}{2\Omega}\right)^4+\cdots\right].
\label{1-T1-1}
\end{eqnarray}
It can be seen that the spin-flip rate is highly sensitive to the effective $g$-factor of the electron, Zeeman energy, hybrid orbital frequency and cyclotron frequency of the QDs.

\section{Results and Discussions}\label{results-and-discussions}

In Fig.~\ref{fig1}, we investigate the contributions of the Rashba and the Dresselhaus spin-orbit couplings on the phonon induced spin relaxation rate as a function of magnetic fields in symmetric InAs QDs. Since the strength of the
Dresselhaus spin-orbit coupling is much smaller than the Rashba spin-orbit coupling ($\frac{\alpha_R}{\alpha_D}= 3.2$ at $E = 10^5 V/cm$ (see Eq.~\ref{coefficient-R-D})), only the Rashba spin-orbit coupling has
a major contribution to the phonon induced spin-flip rate. The cusp-like structure is absent (see Fig.~\ref{fig1} (dashed line)) and the spin-flip rate ($1/T_1$) is a monotonic function of magnetic field ($B$) for pure Dresselhaus case
($\alpha_R= 0$). We solve the corresponding eigenvalue problem with Hamiltonian~(\ref{total}) by applying the exact diagonalization procedure and the Finite Element Method,~\cite{comsol,prabhakar09} obtaining the energy levels. The inset plots show the energy difference vs. magnetic field
(Fig.~\ref{fig1}(i)) and effective Land$\acute{e}$ $g$-factor vs. magnetic field (Fig.~\ref{fig1}(ii)). It can be seen that the level crossing point occurs at $B = 3.5 T$  which is the exact location of the accidental degeneracy point in the spin-flip rate either for pure Rashba case ($\alpha_D = 0$) or mixed cases (both $\alpha_R$ and $\alpha_D $ present). Similar results have been discussed in
Refs.~\onlinecite{bulaev05,bulaev05a}  and we consider these results as a bench mark for further investigation of anisotropic orbital effects on the spin-flip rate in QDs.

Fig.~\ref{fig2} explores the anisotropic effects on the spin-flip rate vs. magnetic fields for the electric fields $E=10^4,5\times 10^4,10^5 V/cm$. It can be seen that the enhancement in the spin-flip rate occurs with the increase in electric fields. The accidental degeneracy point in the spin-flip rate is not affected by the electric fields which tells us that it is purely an orbital effect and is independent of the Rashba-Dresselhaus spin-orbit interaction. In Fig.~\ref{fig2}(ii), the accidental degeneracy point is found at the magnetic field $B = 3.5T$. However, this point increases to the larger magnetic field $B = 6.2T$ in Fig. 2(ii). The extension in the B-field tunability of the spin-flip rate mainly occurs due to an increase in the area of the symmetric quantum dots. Note that the area of the quantum dots in Fig.~\ref{fig2}(ii) is $9$ times larger than the dots in Fig.~\ref{fig2}(i). We quantify the influence of the anisotropic effects on the spin-flip rate in Fig.~\ref{fig2}(iii). Here we find that the quenching in the orbital angular momentum~\cite{pryor06,pryor07} enhances the spin-flip rate and reduces the accidental degeneracy point
to lower magnetic fields ($B = 5.85T$) compared to the symmetric quantum dots ($B = 6.2T$). As a reference, in
Fig.~\ref{fig2}(iii), we also plotted the spin-flip rate vs. magnetic fields (shown by dashed-dotted line) for symmetric QDs ($a = b = 3$) at $E = 10^4 V/cm$ and  $\ell_0=20$ nm. Note that the area of the isotropic and anisotropic quantum
dots in Fig.~\ref{fig2}(ii) and Fig.~\ref{fig2}(iii) are held constant. The expression for the level crossing point is given by the condition~\cite{bulaev05,bulaev05a} $\varepsilon^0_{0,0,-}=\varepsilon^0_{0,1,+}$ i.e., $\hbar\left(\omega_+-\omega_-\right)=|g_0|\mu_B B$ (see Eq.~\ref{epsilon-a}). For isotropic QDs ($a=b=1$), the condition for the level crossing point is $\Omega-\omega_c/2=|g_0\mu_BB/\hbar$. It means, when the difference between the hybrid orbital frequency to the half of the cyclotron frequency becomes equal to the Zeeman frequency then the degeneracy appears in the energy spectrum which give the level crossing point and cusp-like structure near the degeneracy in the
spin-flip rate in QDs. If we compare the condition of the level crossing point for the isotropic and anisotropic
QDs, we find that the anisotropic potential reduces the level crossing point to lower magnetic fields if the area of
the symmetric and asymmetric quantum dots is held constant. However, if we increase the area, the level crossing
point extends to the larger magnetic field.

In Fig.~\ref{fig3}, we study anisotropic effects on the phonon induced spin-flip rate vs. QDs radii in InAs QDs. Similar to Fig.~\ref{fig2}(iii), the anisotropic potential enhances the spin-flip rate and reduces the accidental degeneracy point to lower QDs radii at $\ell_0=43 nm$.

Next, we investigate the phonon induced spin relaxation rate in GaAs QDs. In Fig.~\ref{fig4}, we plot the
phonon induced spin-flip rate vs. magnetic fields for both isotropic ($a=b=8$, (solid line)) and anisotropic ($a=1,
b=64$ (dashed line)) GaAs QDs. It can be seen that the cusp-like structure due to the accidental degeneracy can
be manipulated to lower magnetic fields in the phonon induced spin-flip rate with the application of anisotropic
gate potentials.

The contributions of Rashba and Dresselhaus spin-orbit couplings on the phonon induced spin-flip rate vs.
magnetic fields in GaAs QDs are shown in Fig.~\ref{fig5}. The Rashba and Dresselhaus spin-orbit couplings become
equal at very large electric field $E = 3.02 \times 10^6V/cm$ in GaAs QDs. Below this value of electric field, only the
Dresselhaus spin-orbit coupling has a major contribution on the phonon induced spin-flip rate in GaAs QDs.
However, near the level crossing point (for example, $B = 6 T$ in Fig.~\ref{fig5}), the accidental degeneracy appears
due to only the Rashba spin-orbit coupling, which gives the cusp-like structure in the spin-flip rate.

\section{Conclusions}\label{conclusion}

In summary, we have analyzed in detail anisotropy effects on the electron spin relaxation rate in InAs and
GaAs QDs, using realistic parameters. In Fig.~\ref{fig1}, we have shown that only the Rashba spin-orbit coupling has a
major contribution on the phonon induced spin-flip rate in InAs QDs. In Fig.~\ref{fig2},~\ref{fig3} and ~\ref{fig4}, we have shown that a cusp-like structure due to the accidental degeneracy point appears in the phonon induced spin-flip rate and can be manipulated to lower magnetic fields, in addition to lower QDs radii, with the application of anisotropic gate potentials in III-V semiconductor QDs. Also, we have shown that the anisotropic gate potential causes a quenching effect in the orbital angular momentum that enhances the phonon induced spin-flip rate. Finally, in
Fig.~\ref{fig5} for GaAs quantum dots, we have shown that the Dresselhaus spin-orbit coupling has a major contribution
on the spin-flip rate before and after the accidental degeneracy point, and the Rashba spin-orbit coupling has a
contribution near the cusp-like structure. These studies provide important information for the design and control
of electron spin states in QDs for the purposes of building robust electronic devices and developing solid state quantum computers.

\begin{acknowledgments}
This work has been supported by NSERC and CRC programs (Canada) and by
MICINN Grants No. FIS2008-
04921-C02-01 and FIS2011-28838-C02-01 (Spain).
\end{acknowledgments}


%

\end{document}